\documentclass[pdflatex,sn-mathphys-num,iicol]{sn-jnl}

\usepackage{graphicx}%
\usepackage{multirow}%
\usepackage{amsmath,amssymb,amsfonts}%
\usepackage{amsthm}%
\usepackage{mathrsfs}%
\usepackage[title]{appendix}%
\usepackage{xcolor}%
\usepackage{textcomp}%
\usepackage{manyfoot}%
\usepackage{booktabs}%
\usepackage{algorithm}%
\usepackage{algorithmicx}%
\usepackage{algpseudocode}%
\usepackage{listings}%
\usepackage{textgreek}
\usepackage{cuted}
\usepackage{parskip}

\begin{document}

\title[Binning-Independent Bayesian Analysis of Time-Dependent Perturbed Angular Distribution data]{Binning-Independent Bayesian Analysis of Time-Dependent Perturbed Angular Distribution data}

\author[]{\fnm{Franziskus} \sur{von Spee}}\email{franziskus.spee@ijclab.in2p3.fr}

\affil[]{\orgdiv{Universit{\'e} Paris-Saclay}, \orgname{CNRS/IN2P3, IJCLab}, \orgaddress{\city{91405 Orsay}, \country{France}}}

\abstract{We propose a new approach for analyzing Time-Dependent Perturbed Angular Distribution data. In this, a likelihood is constructed with the help of conditional probabilities of single events and binning of time data is avoided. This likelihood is used in a Bayesian framework to calculate the posterior probability density function for the $g$ factor of interest. This approach is compared to more traditional approaches that use binned data by analyzing simulated datasets. In many cases the resulting posterior probability densities are observed to be multimodal and the results can thus often not be summarized with a single Gaussian approximation. We find that results from    the new approach are more reliable for low-statistics datasets.}

\maketitle

\section{Introduction}
\label{sec:introduction}
With the continuous development and improvement of rare-isotope beam facilities, the field of nuclear physics is drifting towards the study of more and more exotic systems. Such experiments  provide valuable tests of nuclear theories under extreme conditions and offer unique perspectives on nuclear structure. One recent example is a study of doubly-magic $^{78}$Ni \cite{Tan19} that showed that the shell closures $Z=28$ and $N=50$ are valid even close to the neutron drip line. Other recent studies were able to measure the lifetimes of $^{252}$Rf and $^{104}$Te in the sub-\textmu s regime, the shortest-lived nuclear species known to date \cite{Khu25,Cox26}.

While such ambitious experiments are a powerful tool to better understand nuclear structure, they often result in datasets with limited statistics. The aforementioned studies on $^{252}$Rf \cite{Khu25} and $^{104}$Te \cite{Cox26}, e.g., determine lifetime values from less than ten measured events! It is thus an important task to employ analysis methods that are able to extract reliable results and uncertainties from such datasets. For lifetime measurements, an approach that uses Bayesian statistics has been proposed to obtain more reliable uncertainty estimates \cite{Hou21}. Interestingly, this approach also does not require binning of time data.

The gyromagnetic factor -- usually abbreviated to $g$ factor -- of a nuclear state is a property that is often challenging to measure, but provides key insights into the nuclear structure. In many cases, experimental $g$ factors allow to quantify different contributions to the nuclear wave function. Some experimental methods to measure $g$ factors have recently been summarized in Ref. \cite{Geo26}. One of these methods is the measurement of Time Dependent Perturbed Angular Distributions (TDPAD) which is suited for measuring $g$ factors of excited states with lifetimes in the microsecond regime. This work proposes a new method to analyze TDPAD datasets that allows to extract reliable information on $g$ factors even in cases of low-statistics datasets.

\section{Time Dependent Perturbed Angular Distribution Experiments}\label{sec:tdpad}

Time Dependent Perturbed Angular Distribution (TDPAD) experiments aim to measure $g$ factors of excited states of nuclei. Therefore, an ensemble of nuclei in the state of interest is placed in an external magnetic field with strength $B$. The nuclei precess in the magnetic field with the Larmor frequency

\begin{align}
    \omega_L= - \frac{gB\mu_N}{\hbar},
\end{align}
where $\mu_N$ is the nuclear magneton. The nuclei in the state of interest will decay via one or more \textgamma-ray transitions. If the ensemble of implanted nuclei is spin aligned, an anisotropic distribution $W(\vartheta)$ of \textgamma\ rays can be observed where $\vartheta$ is the observation angle with respect to the alignment axis. As in Ref. \cite{Are80}, only the case
\begin{align}
\label{eq:ang_dist}
    W(\vartheta)=1+A_2\left(\frac{1}{4}+\frac{3}{4}\cos(2\vartheta)\right)
\end{align}
with the leading anisotropy parameter $A_2$ is treated in this work, but in general, $W(\vartheta)$ can have additional higher order terms.
The alignment axis precesses with the Larmor frequency and the observation angle with respect to the alignment axis becomes $\vartheta=\theta+\omega_Lt$, where $\theta$ is the observation angle with respect to the initial alignment at $t=0$. The angular distribution then becomes
\begin{align}
    W(\theta,t)=1+A_2\left(\frac{1}{4}+\frac{3}{4}\cos(2\theta+2\omega_Lt)\right).
\end{align}
Here, the anisotropy parameter $A_2$ is an effective parameter that can include several contributions. These contributions are not treated separately in this work, since only the effective $A_2$ enters the analysis. For example, $A_2$ can include the anisotropy of the transition $A'_2$, the alignment $B_2$, and an attenuation factor $q$, such that $A_2=qA'_2B_2$.

In some cases, the initial alignment already depends on the $g$ factor. Then it is useful to replace $\theta$ with $\theta'-\alpha(g)$, where $\theta'$ is the angle of the detector to a fixed axis that can be the beam axis, and $\alpha(g)$ is the initial angle of the alignment axis with respect to the same fixed axis that depends on the $g$ factor. In the following, the discussion will be limited to the case that the initial alignment does not depend on the $g$ factor. For more details on experiments where this is not the case, see Refs. \cite{Geo26,Sto26}.

The time dependence of the angular distribution combines with the exponential decay with lifetime $\tau=1/\lambda$  resulting in a time-dependent \textgamma-ray intensity $N(\theta,t)$ observed at an angle $\theta$ with respect to the initial alignment given by

\begin{align}
    N(\theta,t) \propto \exp(-\lambda t) W(\theta,t).
\end{align}

\subsection{Experimental Setup}

\begin{figure}[htb]
    \centering
    \includegraphics[width=0.85\linewidth]{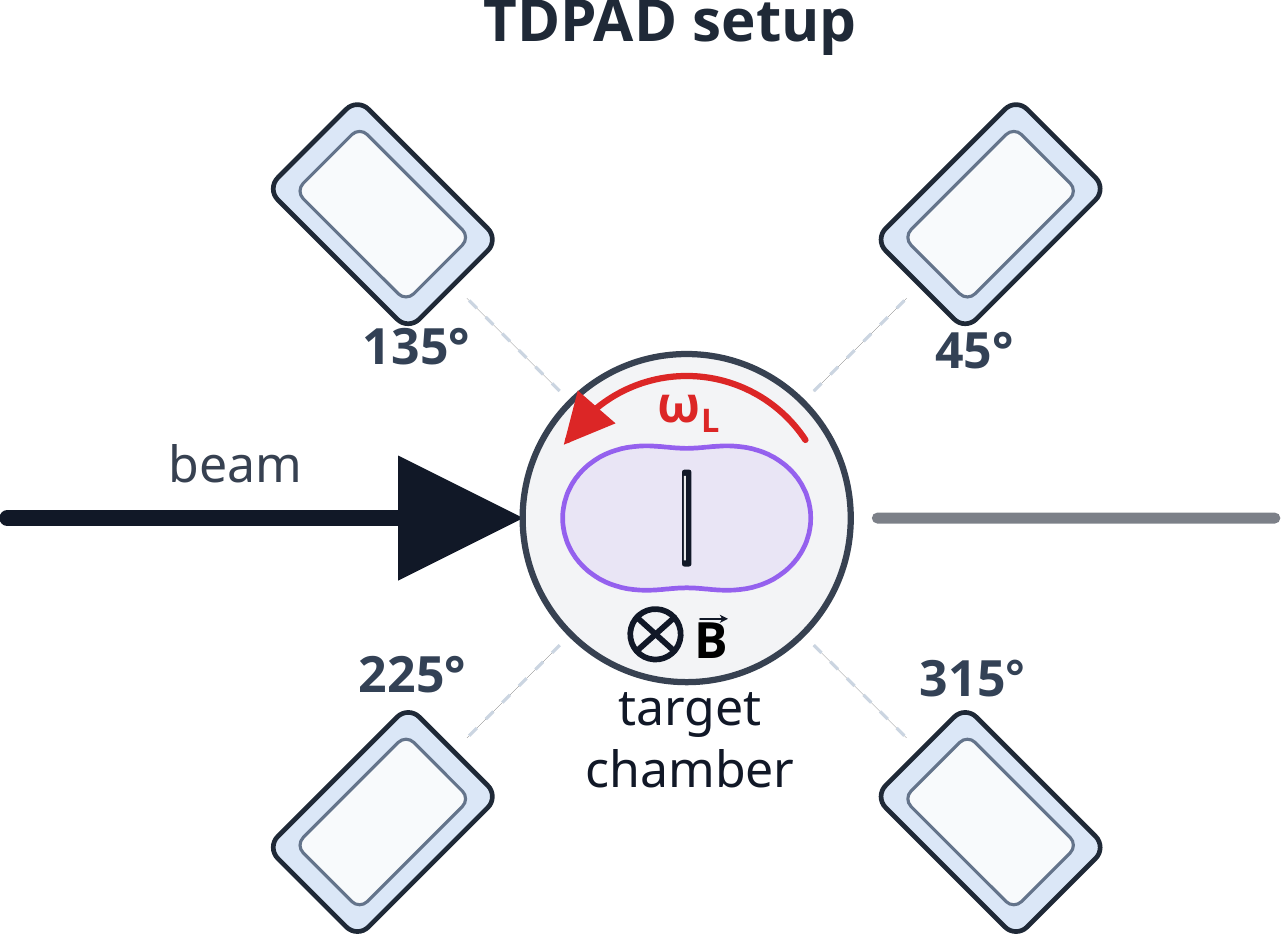}
    \caption{Schematic representation of a typical experimental setup for a TDPAD $g$ factor measurement. Around a target chamber, \textgamma-ray detectors are placed in a plane. A magnetic field is oriented perpendicular to the plane in which the detectors are placed. Target nuclei are implanted in a host, which is located at the center of the target chamber. The nuclei are in an excited state that decays emitting \textgamma\, rays. The angular distribution of the \textgamma\, rays oscillates with the Larmor frequency $\omega_L$.}
    \label{fig:exp_setup}
\end{figure}

To observe the time-dependent angular distributions, \textgamma-ray detectors are placed in the plane orthogonal to the magnetic field around the host where the nuclei of interest are implanted. Typically, four detectors are placed at $90^\circ$ intervals, but more detectors can be added. Due to the symmetry in Eq. (\ref{eq:ang_dist}), two detectors with a relative angle of $180^\circ$ will observe the same time dependent intensity and their data can usually be combined. A schematic drawing of a typical TDPAD setup is presented in Fig. \ref{fig:exp_setup}. In addition, auxiliary detectors are sometimes used that help to clean the data or to determine the implantation time $t_{\mathrm{imp}}$. The \textgamma-ray detectors are labeled with identifiers $i \in \{0,...,I-1\}$, where I is the total number of \textgamma-ray detectors. Each detector has an associated angle $\theta(i)$ with respect to the initial alignment axis and an efficiency $\epsilon(i)$ and observes a time dependent intensity

\begin{align}
    N(i,t)  \propto \ \epsilon(i)  \exp(-\lambda t) \ W(\theta(i),t).
\end{align}

It is also useful to define a ratio function for two detectors $i_0$ and $i_1$

\begin{align}
\label{eq:rt}
    R(t)=\frac{N(i_0,t)/\epsilon(i_0)-N(i_1,t)/\epsilon(i_1)}{N(i_0,t)/\epsilon(i_0)+N(i_1,t)/\epsilon(i_1)}.
\end{align}
In the case of $\theta(i_1)=\theta(i_0)+90^\circ$ and an angular distribution that can be described by Eq. (\ref{eq:ang_dist}), it reduces to

\begin{align}
\label{eq:rt_theo}
    R(t)=\frac{3A_2}{4+A_2}\cos(2\theta(i_0)+2\omega_L t).
\end{align}

\subsection{Alignment methods}
The initial spin alignment of the ensemble of nuclei can typically be achieved in two different ways.
In many cases, the nuclear reaction used to populate the nuclei in the excited state of interest leads to a certain alignment where the alignment axis often coincides with the beam axis. This is typically the case for Coulomb excitation (see, e.g., Ref. \cite{Ald75}) or fusion-evaporation reactions (see, e.g., Ref. \cite{But81}). But also other types of reactions -- for example fragmentation reactions \cite{Ich12} -- can produce spin-aligned ensembles.
A second way of achieving an alignment is by observing a \textgamma-ray transition feeding the state of interest. In this case, the alignment axis is given by the axis between the implantation host and the \textgamma-ray detector that observed the feeding transition (see, e.g., Ref. \cite{Fra65}). If $g$ factors are measured in this way, the method is distinct from a TDPAD and is called Time Dependent Perturbed Angular Correlation (TDPAC) method. The data analysis of TDPAD and TDPAC experiments is, however, sufficiently similar that the analysis method proposed in this work can be applied to both cases. For a more detailed summary of the production of spin-aligned ensembles for TDPAD and TDPAC measurements, see Ref. \cite{Geo26}.

\section{Data Analysis}
\label{sec:data_analysis}
Once an experiment is performed and data are recorded, it is the task of the data analysis to extract the desired information from the data -- which is the information on the $g$ factor in the case of a TDPAD measurement. To this end, first, relevant data are selected from the raw dataset which typically also contains a lot of irrelevant data. Then the relevant data are compared to a model that depends on the $g$ factor.
\subsection{Data selection and preparation}
\begin{figure}[htb!]
    \centering
    \includegraphics[width=0.95\linewidth]{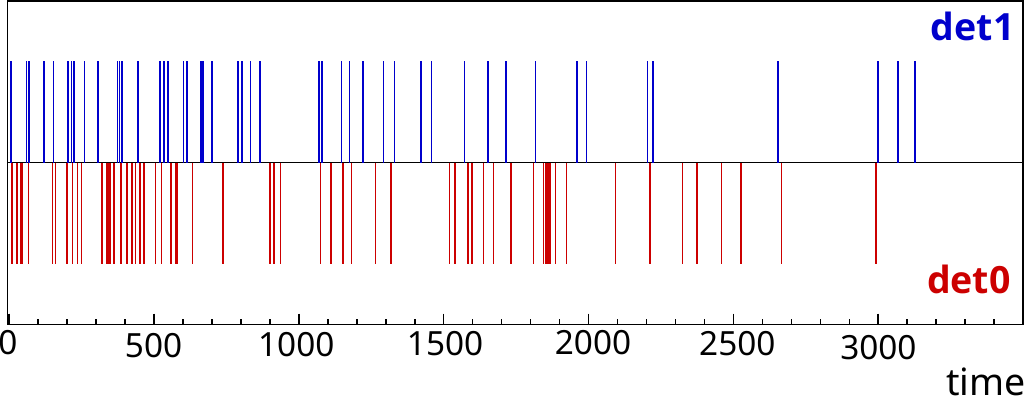}
    \caption{Visualization of a dataset $\mathcal D$ for two detectors with identifiers 0 and 1. Each vertical line represents an event recorded at time $t$. Red vertical lines below the x axis are recorded in detector 0, blue vertical lines above the x axis are recorded in detector 1.}
    \label{fig:dataset_D}
\end{figure}
The recorded data are typically transformed into events of the form $(i,t_s,\mathbf{a})$, where $i$ is a detector identifier, $t_s$ is a timestamp and $\mathbf{a}$ is additional information that might contain coincident events or energy information and is usually used to identify the \textgamma-rays of interest and to transform the raw timestamps $t_s$ into the time after implantation $t=t_s-t_{\mathrm{imp}}$, i.e. the time that the nucleus precesses before emitting the detected \textgamma\ ray. Using the additional information one arrives at a dataset
\begin{align}
\mathcal{D}=\{(i_k,t_k)\}_{k=0}^{K-1}
\end{align}
that contains $K$ events of the form $(i,t)$. Such a dataset is visualized in Fig. \ref{fig:dataset_D} for two detectors with identifiers 0 and 1.
\begin{figure}[htb!]
    \centering
    \includegraphics[width=0.95\linewidth]{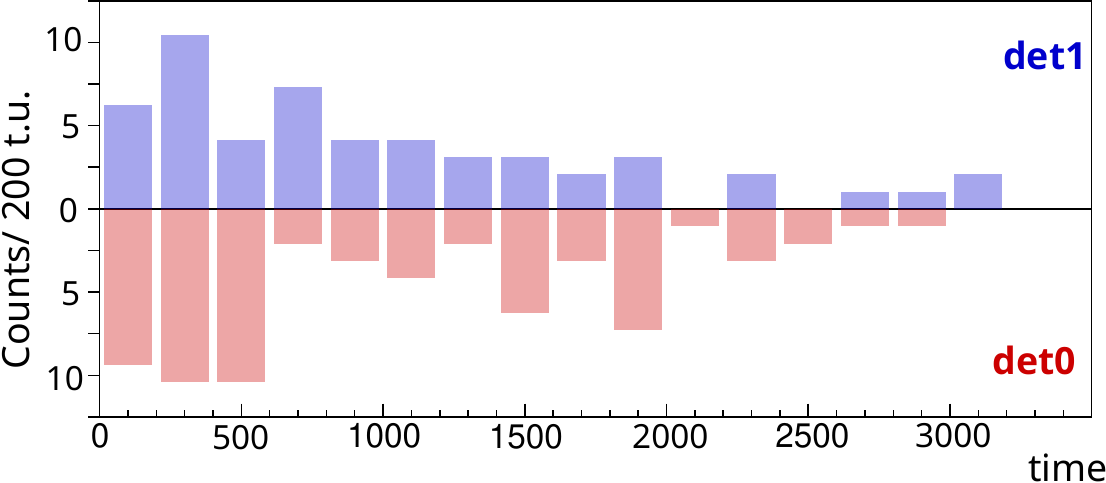}
    \caption{Visualization of the dataset $\mathcal D'$ for two detectors with identifiers 0 and 1 that is obtained by binning the dataset $\mathcal D$ represented in Fig. \ref{fig:dataset_D}.}
    \label{fig:dataset_D'}
\end{figure}

\begin{figure}[htb!]
    \centering
    \includegraphics[width=0.95\linewidth]{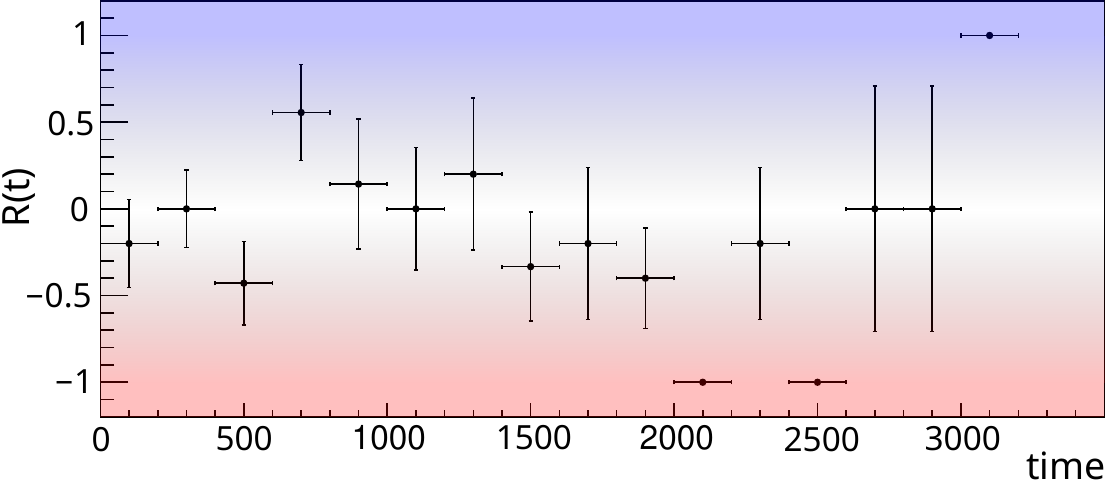}
    \caption{Visualization of the dataset $\mathcal D''$ that is obtained by calculating the ratio function of the dataset presented in Fig. \ref{fig:dataset_D'}.}
    \label{fig:dataset_D''}
\end{figure}

Commonly, the dataset $\mathcal{D}$ is then binned with a certain bin width $\delta t$ and the $K$ events are sorted into $\tilde K$ bins. This can be expressed in the new dataset
\begin{align}
    \mathcal{D}'=\{(i_{\tilde k},\tilde{t}_{\tilde k},N_{\tilde k},\Delta N_{\tilde k})\}_{\tilde k=0}^{\tilde K-1},
\end{align}
where $N_{\tilde k}$ is the number of counts $(i_k,t_k)$ with $\tilde t_{\tilde k} - \delta t /2 < t_k < \tilde t_{\tilde k} + \delta t/2$ and $i_k=i_{\tilde k}$ for a given detector $i_{\tilde k}$ (see Fig. \ref{fig:dataset_D'}). The uncertainty $\Delta N_{\tilde k}$ is in Gaussian approximation given by $\sqrt{N_{\tilde k}}$. Finally, in the case of two detectors a dataset based on the $R(t)$ values defined in Eq. (\ref{eq:rt}) can be constructed:

\begin{align}
    \mathcal{D}''=\{(\tilde{t}_{\tilde k},R_{\tilde k},\Delta R_{\tilde k})\}^{\tilde K -1}_{\tilde k = 0},
\end{align}
 where $\Delta R$ is determined by correctly propagating the uncertainties $\Delta N(\tilde t)$ (see Fig. \ref{fig:dataset_D''}).
 
\subsection{Conventional binned analysis}
\label{sec:Likelihoods}

The datasets $ \mathcal D$, $ \mathcal D'$, and $ \mathcal D''$ all contain information on the $g$ factor. Conventionally, the dataset $ \mathcal D''$ was used to fit the $R(t)$ function defined in Eq. (\ref{eq:rt_theo}) using $\chi^2$ methods with 

\begin{align}
\label{eq:chi2}
    \chi^2_{\mathcal D''}(\phi) = \sum_{\tilde k=0}^{\tilde K-1} \frac{(R_{\tilde k}-R_{\mathrm{theo}}(\tilde t_{\tilde k},\phi))^2}{\Delta R_{\tilde k}^2}.
\end{align}
The parameter tuple $\phi$ contains all free parameters used for the fit. An example would be $\phi=\{g,A_2\}$, but $\phi$ might also contain detector efficiencies, or an unknown initial phase, for example. A prediction of the true parameters is then found at $\chi^2_{\mathrm{min}}$ and the uncertainties using parameters corresponding to $\chi^2_{\mathrm{min}}+1$. This approach works well in cases of sufficient statistics and is well documented, e.g., in Ref \cite{Are80}.
It is also possible to define a related likelihood
\begin{align}
\label{eq:binned_likelihood}
    \mathcal L(\phi|\mathcal D'') \propto \exp\left(-\frac{1}{2}\chi^2_{\mathcal D''}(\phi)\right)
\end{align}
that will have a maximum at $\chi^2_{\mathrm{min}}$ (see \cite{Siv06}).
In cases with low statistics, the need to bin the data for this analysis can lead to problems. On the one hand, every bin should contain a sensible amount of counts, which can be guaranteed by choosing larger bin widths. A larger bin width on the other hand leads to an increased loss of information, in a worst case the criterion $\delta t \ll \pi/\omega_L$ is no longer satisfied and the data become insensitive to the $g$~factor.

\subsection{Binning-independent likelihood}
\begin{figure*}[h!]
    \centering
    \includegraphics[width=0.75\linewidth]{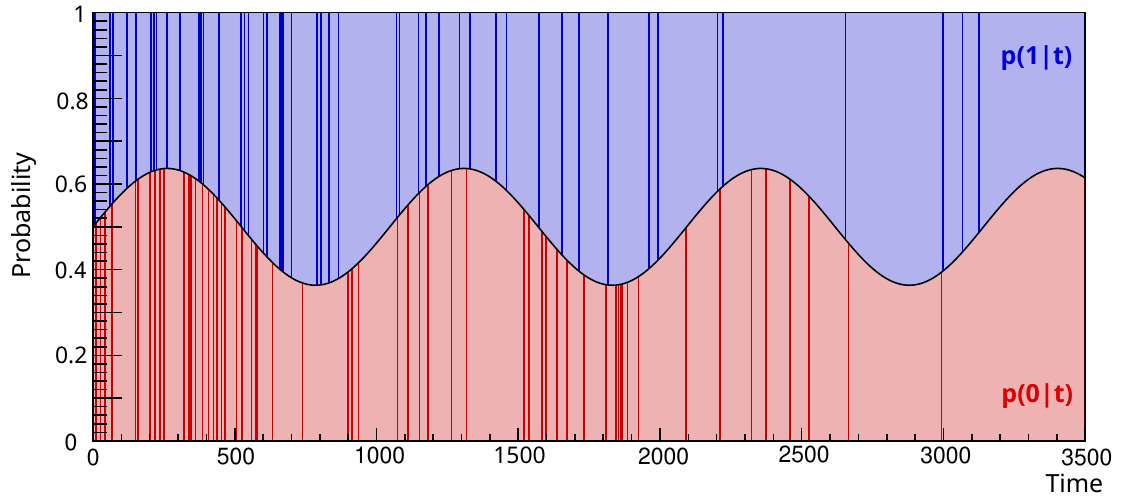}
    \caption{Visualization of the likelihood calculation for two detectors 0 and 1. The red area under the curve symbolizes the probability of an event in detector 0 given time $t$ and the blue area above the curve the probability of an event in detector 1 given time $t$. Note that $p(0|t)+p(1|t)=1\, \forall\, t$. The vertical lines represent the data $\mathcal D$ recorded in detectors 0 or 1, i.e., the realized probabilities. The likelihood is calculated by multiplying all realized probabilities, i.e., the length of all blue and red vertical lines. Thus, the likelihood depends on the form of the curve that is given by the model.}
    \label{fig:vis_likelihood}
\end{figure*}
In order to avoid this problem that arises from binning, in the following an analysis is proposed that uses the unbinned dataset.

To perform an analysis using the unbinned dataset $\mathcal D$, an expression for the probability of a single observation is needed. A single observation is the pair $(i_k,t_k)$ and the probability for this single observation at time $t_k$ in the detector with identifier $i_k$ is given as

\begin{align}
\label{eq:signal_rate}
    p(i_k,t_k) \propto \epsilon(i_k) \exp(-\lambda t_k) W\left(\theta(i_k),t_k\right)
\end{align}
with the normalization condition
\begin{align}
    \sum_{i_k=0}^{I-1} \int_{t_0}^{t_w} p(i_k,t_k) \,dt_k=1
\end{align}
and from this probability one could construct a likelihood. Here $t_0$ and $t_w$ define the observation window. However, in practice, this definition has two disadvantages. First, the integral in the normalization condition is difficult to compute but must be precisely evaluated. In addition, the likelihood depends on the lifetime $\tau=1/\lambda$ of the state of interest.
Instead, it is possible to compute the conditional probability $p(i_k|t_k)$ for a single observation in detector $i_k$ not \textit{at} time $t_k$, but \textit{given} time $t_k$. This changes the normalization. Since the observed time $t_k$ is now a condition and not a possible outcome, the integration over time is omitted and

\begin{align}
    \sum_{i_k=0}^{I-1} p(i_k|t_k)=1.
\end{align}

The probability of a single observation then becomes
\begin{align}
    p(i_k|t_k)=\frac{\epsilon(i_k) \  W\left(\theta(i_k),t_k\right)}
    {\sum_{j=0}^{I-1} \epsilon(j) \  W\left(\theta(j),t_k\right)},
\end{align}
which is independent of the lifetime. The likelihood is then given by multiplying the probabilities of all single observations

\begin{align}
    \mathcal L(\phi|\mathcal D) = \prod_{k=0}^{K-1} p(i_k|t_k),
\end{align}
where the parameter tuple $\phi$ incorporates all free parameters, as in Eqs. (\ref{eq:chi2}) and (\ref{eq:binned_likelihood}). A visualization of the construction and calculation of such a likelihood for a dataset with two detectors is shown in Fig. \ref{fig:vis_likelihood} using the dataset shown in Fig. \ref{fig:dataset_D}.

\subsection{Compton Background Modeling}
\begin{figure}[htb!]
    \centering
    \includegraphics[width=0.95\linewidth]{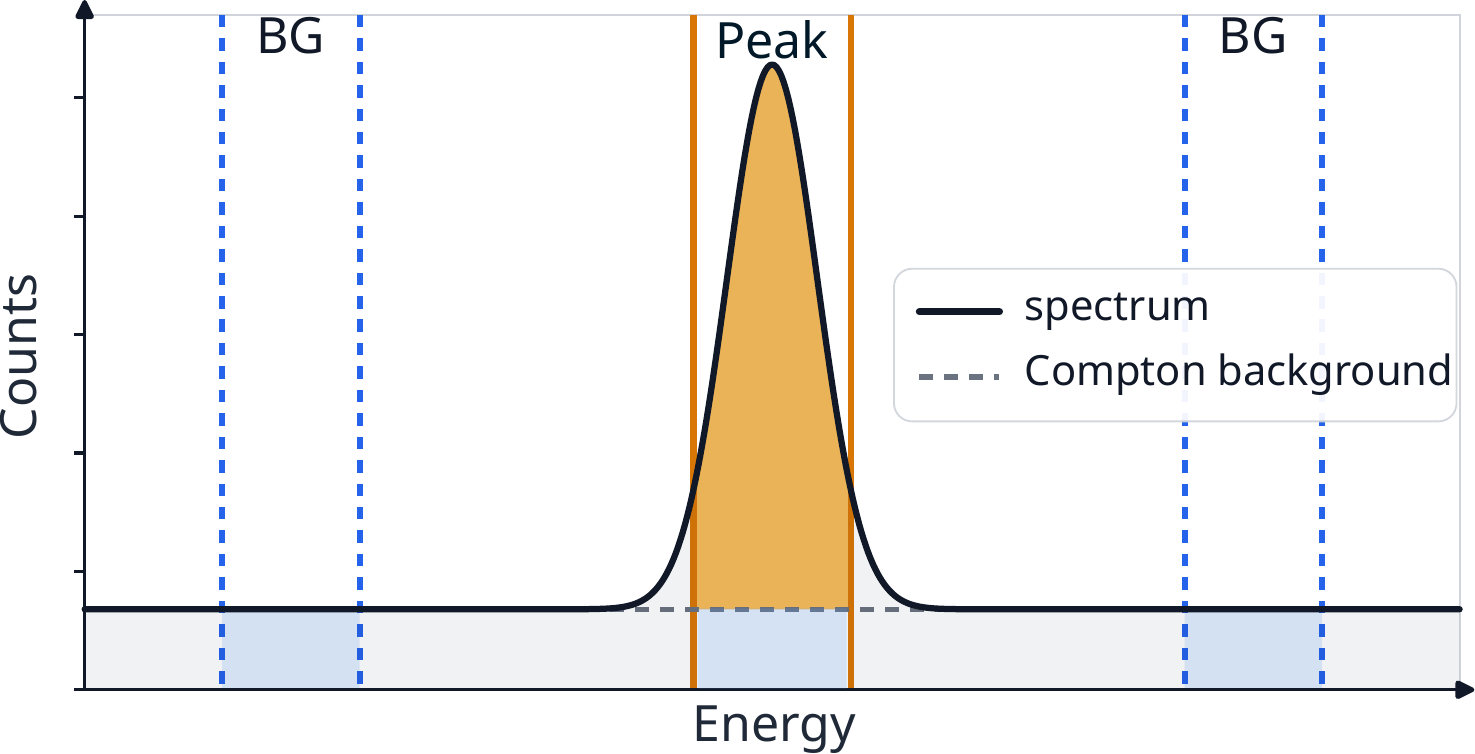}
    \caption{Energy conditions used to obtain signal channel data ($s=1$) and background channel data ($s=0$). The signal channel is selected by the peak gate (solid orange lines), the background channel is selected by the background gate (dashed blue lines). The signal channel will contain a number of background events (blue area) that individually cannot be distinguished from signal data (orange area).}
    \label{fig:compton}
\end{figure}

In many experiments, the signal events are selected via an energy gate and the data will often contain some events stemming from the Compton background (see Fig. \ref{fig:compton}).
Assuming that the time-dependent background events are recorded with a rate $f_B(i,t)$ the single observation probability becomes
\begin{align}
\begin{split}
    p_{S}(i|t)& \propto r\,f_{B}(i,t)+(1-r) f(i,t), \text{\ \ \  with}\\
    f(i,t)&=\epsilon(i) \exp(-\lambda t) W\left(\theta(i),t\right),
\end{split}
\end{align}
where $r \in [0,1]$ is the background-to-total ratio. With the assumption that the Compton background changes only slowly with energy, it is often possible to obtain a background dataset $\mathcal D_B = \{(i_k,t_k)\}_{k=0}^{K_B-1}$ using a background energy gate as depicted in Fig. \ref{fig:compton}. We denote with $w_S$ the relative energy width of the peak gate and with $w_B$ the relative energy width of the background gate. For the background dataset, the single observation probability is then given by
\begin{align}
\begin{split}
    p_B(i|t)&\propto f_B(i,t), \text{ with} \\
    \sum_i p_B(i|t)&=1
\end{split}
\end{align}

The two datasets $\mathcal D$ and $\mathcal D_B$ can be combined to form the dataset
\begin{align}
    \mathcal D_{SB}=\{(i_k,s_k,t_k)\}^{K+K_B-1}_{k=0},
\end{align}
where $s$ is a flag with $s=0$ for the background data channel and $s=1$ for signal data channel that also contains background data. The probability for a single event in detector $i$ and channel $s$ given time $t$ is then given by
\begin{align}
\begin{split}
    &p_{SB}(i,s|t) \propto \\
    &
    \begin{cases}
        w_Br\,f_B(i,t) &\text{if } s=0 \\
        w_S \left(r\,f_B(i,t) + (1-r) f(i,t)\right) &\text{if } s=1       
    \end{cases}
\end{split}
\end{align}
and the normalization condition
\begin{align}
    \sum_{s=0}^1\sum_{i=0}^{I-1}p_{SB}(i,s|t) =1.
\end{align}
Here, $w_B$ and $w_S$ function as weights that are necessary since the background data channel and the signal data channel are usually obtained with energy gates of different sizes.
The Compton background rate $f_B(i,t)$ is in general a composite structure. Assuming that the Compton background is mostly isotropic, the Compton background rate can often be modeled as
\begin{align}
    f_B(i,t) = \epsilon(i) \sum_b c_b\exp(-\lambda_b t).    
\end{align}

\begin{figure*}[htb!]
    \centering
    \includegraphics[width=0.75\linewidth]{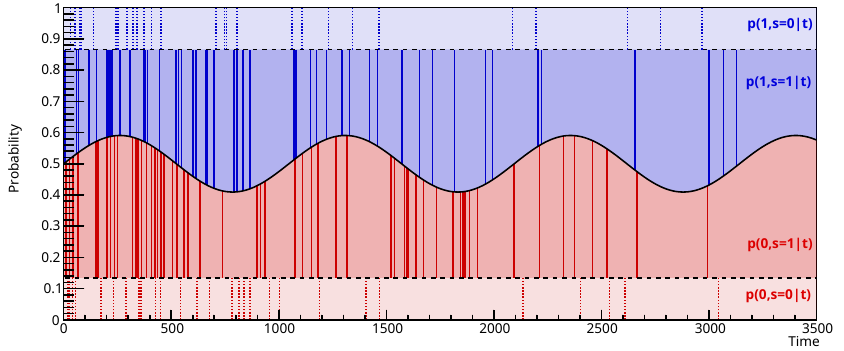}
    \caption{Visualization of the likelihood calculation for two detectors 0 and 1. For both detectors the background is non-negligible and a background channel $s=0$ is taken into account while calculating the likelihood. The background is assumed to have the same lifetime as the signal resulting in a constant probability to observe a background event. Realized probabilities are marked with solid vertical lines for the signal channel and with dashed lines for the background channel. The likelihood is the product of all realized probabilities including signal and background channels.}
    \label{fig:likelihood_4channels}
\end{figure*}
For different time regimes, the Compton background is dominated by different lifetimes. For small $t$, contributions with small lifetimes $1/\lambda_b$ dominate, for large $t$, contributions with large lifetimes dominate. How the Compton background can be approximated depends on the exact case. In many cases, it is possible to find a region in time where the background rate reduces to
\begin{align}
\label{eq:bg_rate}
    f_B(i,t) = \epsilon(i) \exp(-\lambda_B t).      
\end{align}
The single observation probability $p_{SB}(i_k,s_k|t_k)$ then depends only on the difference between the involved decay constants $\lambda-\lambda_B=\Delta\lambda$ that can be treated as a single parameter. This parameter disappears if $\lambda\approx\lambda_B$. With a model for the background rate, the single observation probability can be calculated. Assuming a background rate as in Eq. (\ref{eq:bg_rate}) the single observation probability becomes


\begin{align}
\begin{split}
p_{SB}(i_k,s_k\mid t_k)
={}&
\frac{\bigl((1-s_k)w_B+s_kw_S\bigr)r\epsilon(i_k)e^{\Delta\lambda t_k}}{T(t_k)}\\
+&\frac{s_kw_S(1-r)\epsilon(i_k)W\bigl(\theta(i_k),t_k\bigr)}{T(t_k)},
\end{split}
\end{align}
with the denominator
\begin{align*}
\begin{split}
T(t_k)
=
&(w_B+w_S)r e^{\Delta\lambda t_k}\sum_{j=0}^{I-1}\epsilon(j)\\
+&w_S(1-r)\sum_{j=0}^{I-1}\epsilon(j)W\bigl(\theta(j),t_k\bigr).
\end{split}
\label{eq:psb}
\end{align*}

This expression looks clunky but allows a straightforward calculation of the likelihood that depends on the parameter tuple $\phi=\left(g,A_2,r,\Delta\lambda\right)$ if the efficiencies are known. A visualization of such a calculation of the likelihood is displayed in Fig. \ref{fig:likelihood_4channels}. Similar expressions can be found if the background rate $f_B(i,t)$ takes a different form.

Here, the discussion of background modeling is limited to Compton background. There are many other possible sources of background, such as random coincidences, atomic flash, and many others. A full discussion of all possible backgrounds is beyond the scope of this work. It will, e.g., not always be possible to obtain information on the background from a background energy gate adjacent to the signal energy gate. How exactly the background should be modeled and how information on the background can be obtained depends on each individual experiment.

\subsection{Bayesian inference}\label{sec6}
In the last sections, different expressions for the likelihood were derived that use different approaches and are suited for different experimental situations. In a Bayesian framework these likelihoods are then used in  Bayesian inference to learn something about the parameter of interest $g$. The result of a Bayesian inference is the posterior probability density

\begin{align}
    P(\phi|\mathcal D) = \frac{\mathcal L(\phi|\mathcal D) \rho(\phi)}{\mathcal E (\mathcal D)},
\end{align}
where $\rho(\phi)$ is the prior of the parameter tuple $\phi$ and $\mathcal E(\mathcal D)$ is the evidence. The latter is often difficult to compute and in many cases only a quantity proportional to the posterior is calculated, which is usually sufficient. The posterior $P(\phi|\mathcal D)$ contains information on all model parameters $\phi$. Usually, only a limited number of parameters are of interest, e.g. $g$ and $A_2$ or even only $g$. To obtain information only on this limited number of parameters, the posterior is marginalized, i.e. an integration over the parameter tuple $\phi_n$ containing all other parameters is performed:
\begin{align}
    P(g|\mathcal D) = \int P(\phi|\mathcal D) d\phi_n,
\end{align}
with $\phi=\left(g,\phi_n\right)$.

\subsubsection{Choice of prior}
In Bayesian inference, the prior $\rho(\phi)$ reflects the knowledge about the parameters $\phi$ before the experiment. Prior knowledge about physical parameters like $g$ and $A_2$ comes from the understanding of the physics. The Schmidt limits constrain the expected $g$ factor values \cite{Sch37}. The expected $A_2$ parameter is governed by the alignment mechanism. In the case of a TDPAC measurement with known involved spins and multipolarities for example, the theoretical $A_2$ parameters are well known, but might be attenuated by less known geometrical effects \cite{Ros53}. For physical parameters, it is also possible to use the posterior of another already analyzed experiment as a prior. This sequential Bayesian updating is one of the advantages of Bayesian inference. The priors of experimental parameters like detector efficiencies and detector angles usually stem from calibration measurements or the geometry of the setup. For a more detailed discussion on priors in Bayesian inference, the reader is referred to, e.g., Ref. \cite{Siv06}.

\subsubsection{Summarizing posteriors}
In benign cases with sufficient statistics, the posterior $P(g|\mathcal D)$ can often be approximated by a single Gaussian with mean $\mu$ and variance $\sigma^2$ which allows a summary of the data analysis using a single value with a single uncertainty. Such a summary is desirable, since this form is commonly used for experimental results and intuitively understood. In cases of data with limited statistics, however, Gaussian approximations of posteriors often do not work. Then, a summary of the result with a single value and a single uncertainty is insufficient if not misleading (cmp. \cite{Siv06}). Another way to summarize a posterior is the designation of regions with the highest probability density (HPD) such that the integral of the regions adds up to 68\% or 95\%. This summary also works in cases where the posterior cannot be approximated as a Gaussian but will give a close to standard summary if the posterior can be approximated as a Gaussian. 

\section{Test on simulated data}
\begin{figure*}[htb!]
    \centering
    \includegraphics[width=0.95\linewidth]{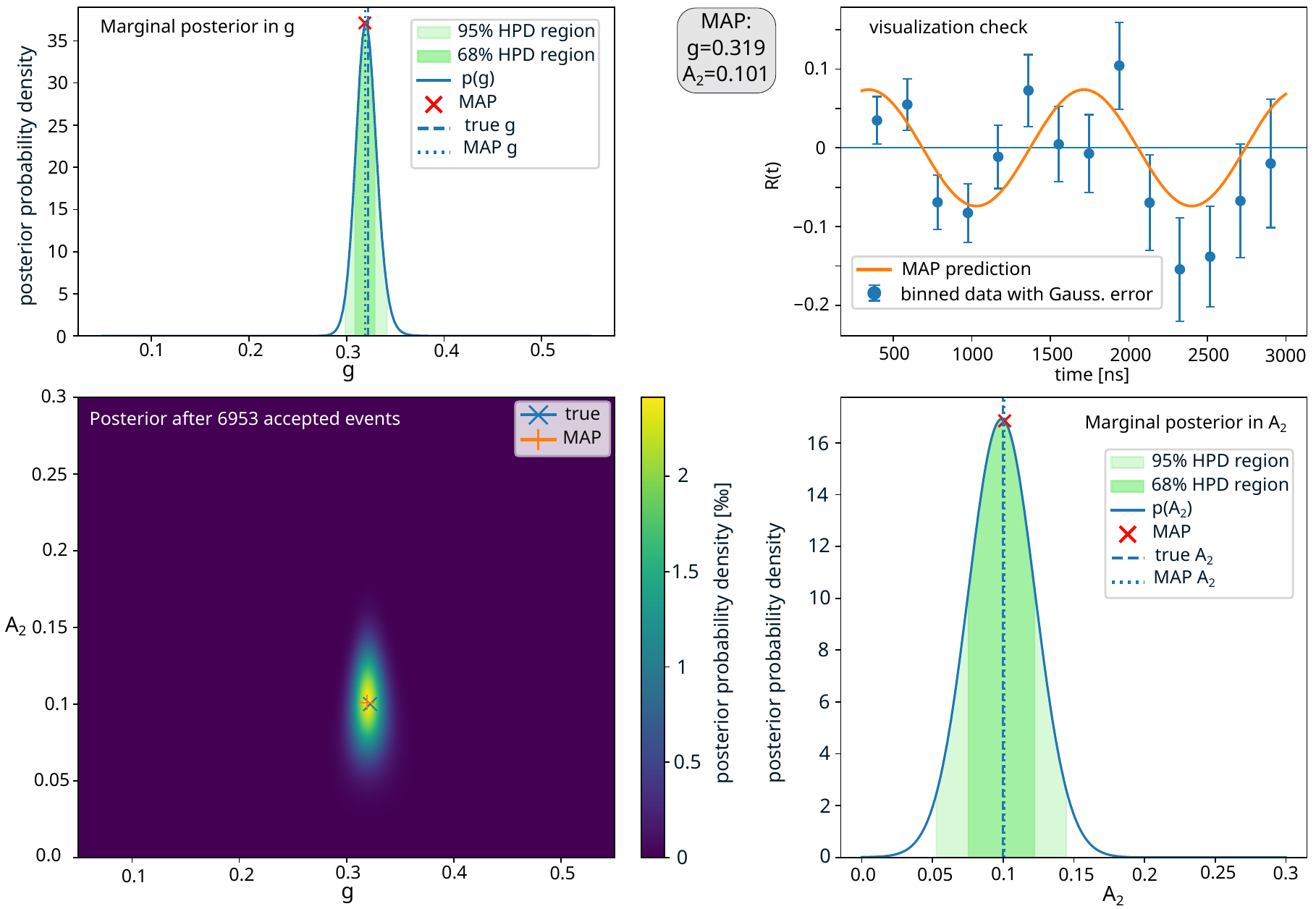}
    \caption{Result of the binning-independent analysis of a simulated dataset with 10\,000 events of which 6953 lie in the observation window. The plot on the bottom left shows the full posterior probability distribution, marginalizations in $g$ and $A_2$ are shown on the top left and bottom right, respectively. The top right plot shows the prediction of the $R(t)$ function using the parameters of the MAP together with the data that are binned with a bin width of 193 ns. For details, see text.}
    \label{fig:simul_10000}
\end{figure*}
To test the different analysis procedures and compare their results, we have developed a tool to generate TDPAD datasets and analyze them. For this, a setup with two detectors that have the same efficiency is assumed. An aligned ensemble of nuclei in the state of interest is simulated. Adjustable parameters of the tool are the detector angles $\theta_{0,1}$, the external magnetic field strength $B$, the lifetime $\tau$ and $g$ factor of the state of interest, as well as the alignment parameter $A_2$. With these parameters chosen, a fixed number of data points of the form $(i,t)$ are generated with a probability given by Eq. (\ref{eq:signal_rate}) to form a dataset. It is assumed that the dataset is free of background.

For the analysis, only data points within a chosen range $[t_0,t_w]$ are considered and the Posterior is calculated using both the binning-independent and the binned likelihood. The prior is flat in a region defined by $[A_2^{\mathrm{min}},A_2^{\mathrm{max}}]$ and $[g^{\mathrm{min}},g^{\mathrm{max}}]$.
In a first test, a dataset with 10\,000 events was simulated using $A_2=0.1$, $g=0.322$ and $\tau=1300$ ns. Detectors were placed at $45^\circ$ and $135^\circ$ and a magnetic field of $0.15$ T was assumed. The result of the analysis using the binning-independent likelihood is shown in Fig. \ref{fig:simul_10000}. In this case of relatively high statistics, the posterior has an approximately Gaussian form in both $A_2$ and $g$. The maximum of the posterior (usually called maximum a posteriori or MAP) lies close to the true value and the true value is contained in the $68\%$ HPD region.
\begin{figure*}[htb!]
    \centering
    \includegraphics[width=0.95\linewidth]{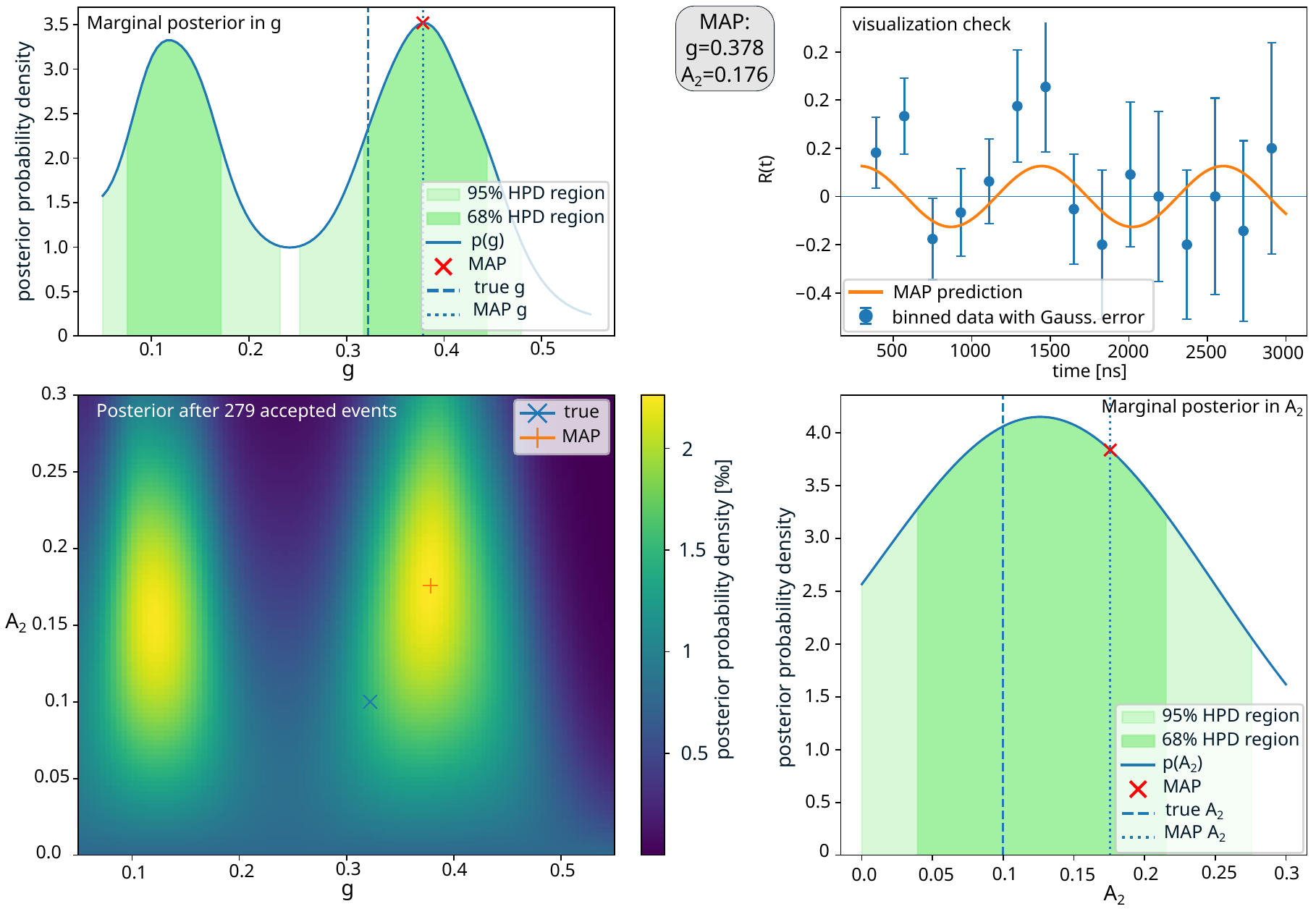}
    \caption{Result of the binning-independent analysis of a dataset of 400 events of which 279 lie in the observation window. These data are a subset of the data analyzed in Fig. \ref{fig:simul_10000}. The plots are analogous to the plots shown in Fig. \ref{fig:simul_10000}. A bin width of 180 ns was chosen for the visualization on the top right. For details, see text.}
    \label{fig:simul_400}
\end{figure*}

In a second test, a subset of 400 events was drawn from the same dataset and analyzed. Figure \ref{fig:simul_400} shows the result of the analysis of this low-statistics dataset. The posterior is in this case multimodal in $g$ and cannot be described by a single Gaussian. However, the posterior does already contain information on the $g$ factor: The true value lies in the $68\%$ HPD region.

This multimodal behavior of the marginalized posterior in $g$ was observed in nearly all analyses of simulated datasets with limited statistics. It is important to stress that this does not mean that there are multiple frequencies present in the dataset. The dataset was simulated using only one $g$ factor. But with a limited level of statistics, the analysis is not yet able to clearly identify this one frequency and in some cases more than one frequency seems plausible, resulting in multimodal posteriors.
\begin{figure*}
    \centering
    \includegraphics[width=\linewidth]{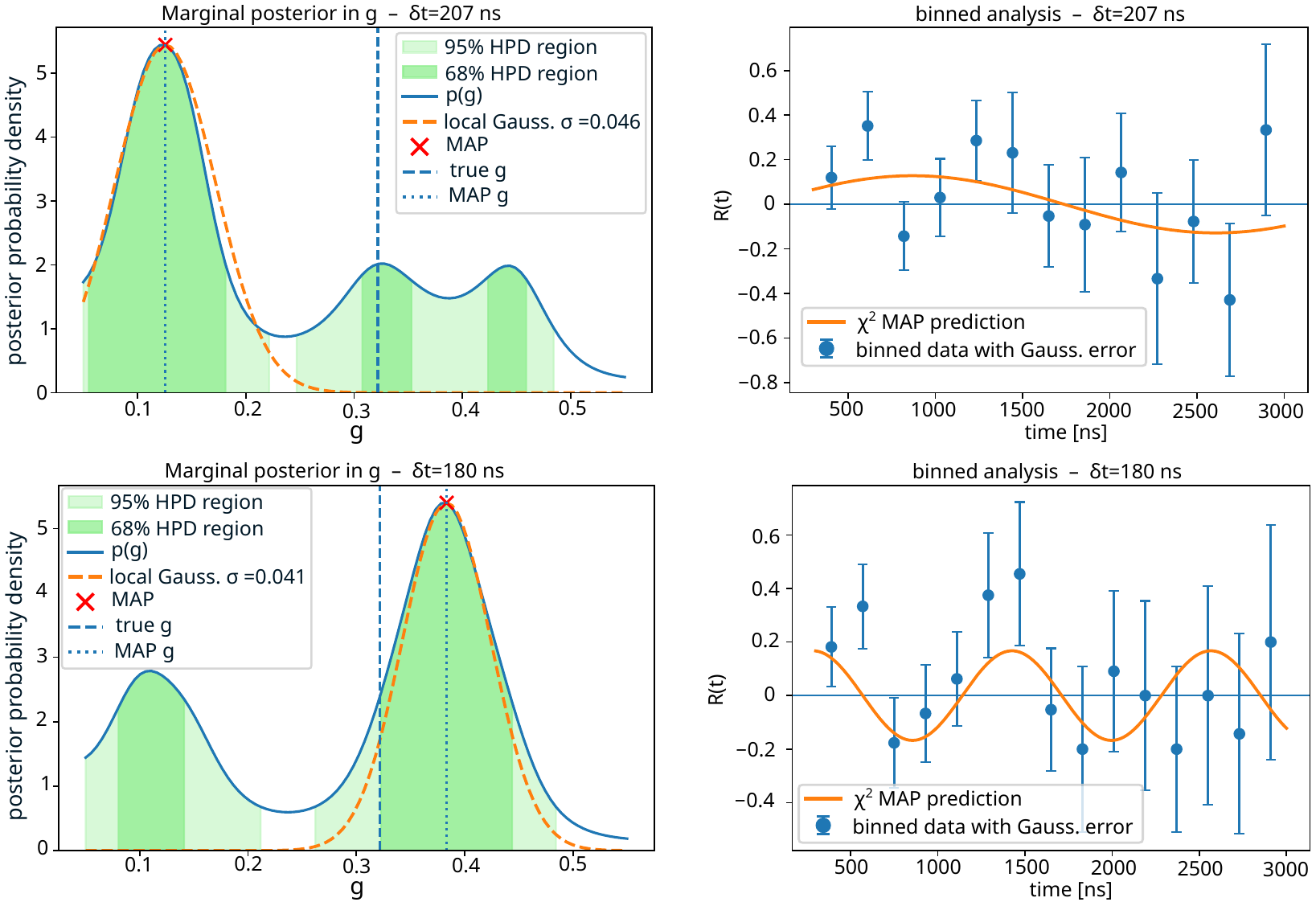}
    \caption{Result of the binning-dependent analysis a dataset of 400 events of which 279 lie in the observation window. These data are the same data that are analyzed in Fig. \ref{fig:simul_400}. The results using a bin width of 207 ns (top) are compared to the results using a  bin width of 180 ns (bottom). In addition, the Gaussian approximation of the posterior in $g$ around the MAP is shown.}
    \label{fig:bin_comparison}
\end{figure*}

So far we have looked at the results of analyses using binning-independent likelihoods. Now we can compare these results with the results of a likelihood that uses binned data as in Eq. (\ref{eq:binned_likelihood}). A problem with this comparison is that the likelihood that uses binned data will depend to some degree on the chosen bin width. Results of such an analysis for two different bin width choices are shown in Fig. \ref{fig:bin_comparison}. Notably, while the posteriors are relatively similar to each other and also to the posterior shown in Fig. \ref{fig:simul_400}, in one case the true value lies in the $68\%$ HPD region, in the other it does not.
If we approximate the results of the binned analysis with a Gaussian through the MAP, we obtain a result that is equivalent to the result of a classical $\chi^2$ analysis, where the uncertainties are determined by $\chi_{\mathrm{min}}^2+1$. In Fig. \ref{fig:bin_comparison}, such Gaussian approximations are shown. However, in both cases, the Gaussian approximation around the MAP is not justified and would underestimate the probability for the true $g$ factor value.
\begin{figure*}
    \centering
    \includegraphics[width=0.99\linewidth]{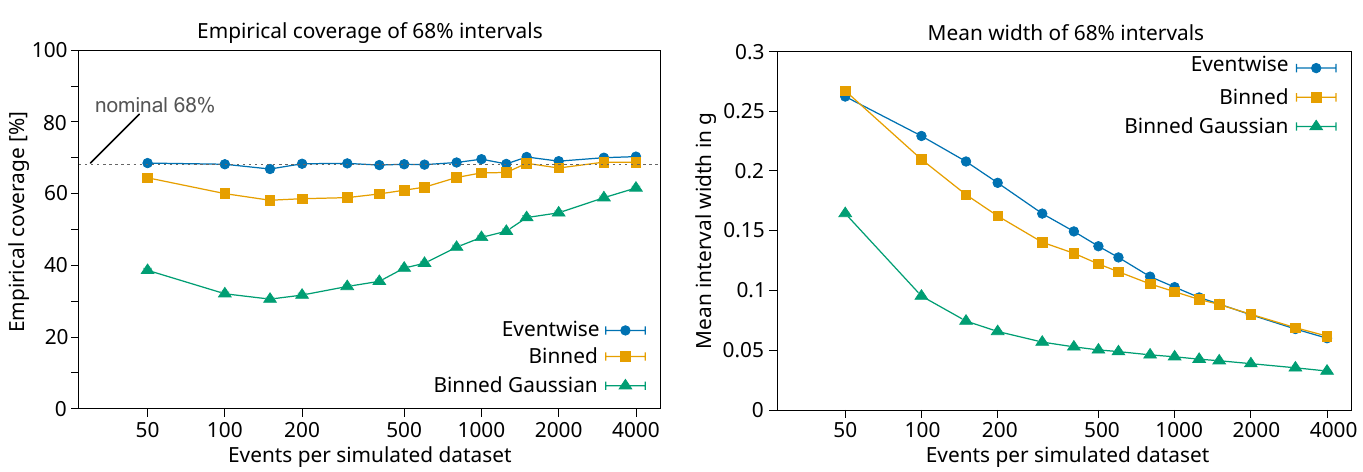}
    \caption{Results of the coverage study comparing the three different analysis methods. In blue circles, the results using the binning-independent likelihood are shown. The orange squares represent the results of the binned analysis and green triangles the results of the Gaussian approximation. On the left-hand side, the empirical coverage of the $68\%$ HPD region is shown for different levels of statistics. On the right-hand side, the average width of the nominal $68\%$ HPD region is plotted depending on the level of statistics.}
    \label{fig:cov_study}
\end{figure*}

A coverage study was performed to quantify the reliability and predictive power of three different analysis approaches. The first approach uses the binning-independent likelihood, the second the binned likelihood and the third a Gaussian approximation of the binned likelihood. For the coverage study, a scenario with known lifetime but unknown $g$ factor and $A_2$ parameter was assumed. The lifetime was set to $1300$ ns, the detectors were again placed at $45^\circ$ and $135^\circ$. The observation window was set from 300 to 3000 ns and a magnetic field of $0.15$ T was simulated. Datasets with different levels of statistics between 50 and 4000 events were simulated. For each level of statistics, 10\,000 datasets were simulated. For each of the simulated datasets, a true $g$ factor value and $A_2$ parameter were drawn from a flat prior range of $g\in[0.05,0.55]$ and $A_2\in[0,0.3]$. The same flat prior range was used to analyze the data. Each of the datasets was analyzed using the three different approaches and noting down whether the true $g$ factor value lay in the $68\%$ HPD region and how large the $68\%$ HPD region was. For the binned analysis, a bin width of 225 ns was chosen for all analyses. The results of the coverage study are shown in Fig. \ref{fig:cov_study}. Evidently, the analysis using the binning-independent likelihood gives the most reliable results, even for low levels of statistic. For higher levels of statistics, also a binned likelihood gives reliable results that in practice do not differ significantly from those of a binning-independent analysis. For lower levels of statistic, the approach using a binned likelihood seems to be somewhat overconfident. The Gaussian approximation of the posterior is the least reliable analysis method and is frequently overconfident, even for higher levels of statistic. This said, whether a Gaussian approximation is valid depends on the specific case, and if the form of a specific posterior can be approximated by a Gaussian, there is no reason to not use that approximation. The coverage study just shows that a Gaussian approximation will not be valid in all cases and one should be careful before using one.
To illustrate the statistical fluctuations encountered in such analyses, the simulation and analysis tool used in this section has been made publicly available at \url{https://tdpad-toy.streamlit.app/}.

The tests performed in this section have all assumed the simplest case, where there is no background, and detector positions and efficiencies are known. For more complicated cases, the methods of Bayesian inference become more cumbersome than a simple scan of the parameter space and large coverage studies are thus not feasible. First analyses of datasets with background and unknown efficiencies show, however, that the multimodal character of the posterior in $g$ persists for datasets with limited statistics.

\section{Summary}
In this work, a new method to analyze TDPAD data was introduced. There are two key differences compared with a conventional analysis: First, the new method avoids binning or rebinning of time information. Second, a Bayesian approach is used that is suited for low-statistics scenarios. By calculating a binning-independent likelihood and not using binning-dependent \textchi$^2$-methods, the best available time information is used and no time information is lost. In addition, the choice of the bin width, which often opens the door for subjective bias to enter the analysis process, becomes unnecessary. The posterior resulting from an analysis using Bayesian inference allows results to be quantified that cannot be expressed with a single value and a single uncertainty. This is especially valuable for the analysis of TDPAD datasets with low statistics that were shown to frequently lead to posteriors that are multimodal -- independent of whether a binned or binning-independent method is used. In this context, a tool is provided that allows readers to familiarize themselves with an analysis of low-statistics TDPAD datasets using the proposed method and comparing it to a binned analysis.  A coverage study showed that the new analysis method can extract more reliable information from a dataset than both a conventional approach and a Bayesian approach using binned data. The Bayesian approach also allows for sequential Bayesian updating, where an obtained posterior is used as a prior for a new dataset. This makes it possible to combine the full information of multiple datasets. The presented improvements in the analysis method thus increase the feasibility of future experiments on $g$ factors of excited states of exotic species.

\subsection*{Acknowledgements}
The author would like to thank K. Stoychev, H. v. Campe, C.-D. Lakenbrink and G. Georgiev for helpful advice, fruitful discussions and thorough proofreading.

\subsection*{Data Availability Statement}
This manuscript has no associated data.

\subsection*{Code Availability Statement}
The simulation and analysis tool used for demonstrating the method is made publicly available at \url{https://tdpad-toy.streamlit.app/}. The source code is archived at \url{https://doi.org/10.5281/zenodo.22674809}. A full analysis framework is currently under development at \url{https://github.com/franzspee/Full_BIA}.

\bibliography{sn-bibliography}

\end{document}